\documentclass{emulateapj}
\usepackage{color}
\newcommand{\WDTW}{SDSS J125733.63+542850.5}
\newcommand{\WDT}{SDSS 1257+5428}
 
\newcommand{\sw}{SWARMS}

\slugcomment{Draft Version \today}

\shorttitle{First results from \sw: \WDT}
\shortauthors{Badenes et al.}

\begin{document}

\title{First Results from the \sw\ survey. \WDT: A Nearby, Massive White Dwarf Binary With a Likely Neutron Star or Black Hole Companion.}


\author{Carles Badenes\altaffilmark{1,2,3,4}, Fergal Mullally\altaffilmark{1}, Susan E. Thompson\altaffilmark{5} and Robert H. Lupton\altaffilmark{1}}

\altaffiltext{1}{Department of Astrophysical Sciences, Princeton University. Peyton Hall, Ivy Lane, Princeton NJ
  08544-1001; badenes@astro.princeton.edu, mullally@astro.princeton.edu, rhl@astro.princeton.edu}

\altaffiltext{2}{\textit{Chandra} Fellow}

\altaffiltext{3}{Benoziyo Center for Astrophysics, Faculty of Physics, Weizmann Institute of Science, 76100 Rehovot, Israel}

\altaffiltext{4}{School of Physics and Astronomy, Tel-Aviv University, 69978 Tel-Aviv, Israel}

\altaffiltext{5}{Department of Physics and Astronomy, University of Delaware, Newark, DE 19716; and Delaware
  Asteroseismic Research Center, Mt. Cuba Observatory, Greenville, DE 19807; sthomp@physics.udel.edu}

\begin{abstract}
  We present the first results from the \sw\ survey, an ongoing project to identify compact white dwarf (WD) binaries in
  the spectroscopic catalog of the Sloan Digital Sky Survey. The first object identified by \sw, \WDT, is a single-lined
  spectroscopic binary in a circular orbit with a period of $4.56$ hr and a semiamplitude of
  $322.7\pm6.3\,\mathrm{km\,s^{-1}}$. From the spectrum and photometry, we estimate a WD mass of
  $0.92^{+0.28}_{-0.32}\,\mathrm{M_{\odot}}$. Together with the orbital parameters of the binary, this implies that the
  unseen companion must be more massive than $1.62^{+0.20}_{-0.25}\,\mathrm{M_{\odot}}$, and is in all likelihood either
  a neutron star or a black hole. At an estimated distance of $48^{+10}_{-19}$ pc, this would be the closest known
  stellar remnant of a supernova explosion.
\end{abstract}


\section{INTRODUCTION}
\label{sec:Intro}

The vast majority of stars with masses below $8 \mathrm{M_{\odot}}$ end their lives as white dwarfs (WDs). Left to their
own devices, WDs will just cool and fade away, but WDs that form part of close binary systems can interact with their
companions, steering away from the normal course of isolated stellar evolution and giving rise to a plethora of
astrophysical phenomena. When the companion of the WD is another compact object (a second WD, a neutron star [NS], or a
black hole [BH]), we refer to these systems as compact WD binaries, or CWDBs. The orbital evolution of detached CWDBs is
governed by the emission of gravitational waves
\citep[GWs,][]{peters63:GW_radiation_Keplerian_orbit,paczynski67:GW_close_binaries}. For systems with small enough
separations, loss of angular momentum leads to a merger of the two components in a time that is short enough to be of
astrophysical interest. This kind of CWDBs constitute the largest population of Galactic GW sources
\citep{evans87:WD_binaries_as_GW_sources}, dominating the GW foreground for future missions like LISA
\citep{nelemans09:binaries_review}. They can also provide important constraints on the still poorly understood process
of common-envelope evolution in binary systems \citep{nelemans05:common_envelope_in_WD_binaries}.

But perhaps the most interesting aspect of CWDBs is the outcome of the merging process at the end of their binary
evolution. For systems composed of two CO WDs whose combined mass is above the Chandrasekhar limit, the merger might
trigger a thermonuclear runaway and lead to a Type Ia supernova (SN) explosion
\citep{iben84:typeIsn,webbink84:DDWD_Ia_progenitors}. This possibility, often referred to as the double degenerate WD
(DDWD) SN Ia progenitor scenario, is the only theoretical model that naturally explains the absence of H in the nebular
spectra of Type Ia SN \citep[see][and references therein]{Leonard07:H_nebular_Ia_spectra}, and it has motivated a number
of searches for suitable candidate systems. The most comprehensive of these searches, the SPY survey
\citep{napiwotzki01:SPY_survey}, examined the spectra of $\sim1000$ WDs to look for radial velocity (RV) shifts in the
characteristic absorption lines. The total number of detached DDWDs found to date by SPY and other surveys is around 100
\citep{napiwotzki04:SPY_04}, but the periods and mass estimates have only been published for 24 systems
\citep{nelemans05:SPY_IV}. So far, none of these systems clearly fulfills the requisites to be a SN Ia progenitor.

In this paper and in a companion publication \citep{mullally09:DDWDs}, we present the first results from \sw, the Sloan
White dwArf Radial velocity data Mining Survey. The aim of \sw\ is to mine the WD catalog in the spectroscopic data base
of the Sloan Digital Sky Survey \citep[SDSS,][]{york00:SDSS_Technical} in search of CWDBs. Our ultimate goal is to find
the DDWD progenitors of Type Ia SNe, or at least put constraints on the rate of DD mergers in the Galactic disk that can
be compared with measurements of the local Type Ia SN rate. This will allow us to assess the viability of the DDWD
progenitor scenario for SN Ia. This paper is organized as follows. Our data mining strategy is briefly described and
compared to the SPY survey in Section \ref{sec:Strategy}, using as an example the CWDB \WDTW\ (henceforth, \WDT), the
first object discovered by \sw. In Section \ref{sec:Followup}, we present the follow-up observations of \WDT\ that we
have conducted in order to determine its orbital and spectral parameters. In Section \ref{sec:Companion}, we discuss the
implications of our findings for the nature of the unseen companion of \WDT. We summarize our conclusions and outline
the future prospects for our survey in Section \ref{sec:Conclusions}.

\section{\sw: MINING THE SDSS DATA BASE FOR COMPACT WD BINARIES}
\label{sec:Strategy}

\subsection{General Strategy}

The first version of the SDSS WD catalog was published by \citet{kleinman04:SDSS_WDs}. The most recent version,
presented in \citet{eisenstein06:WD_SDSS_DR4} (henceforth, E06), corresponds to Data Release 4 of the SDSS
\citep{adelman06:SDSS_DR4}, and it is currently the largest catalog of its kind, with 9316 spectroscopically confirmed
WDs. This number should be roughly doubled in the forthcoming DR7 version. Many publications have analyzed the SDSS
spectra of WDs and related objects like subdwarf stars and cataclysmic variables (see E06 for a list of references), but
so far all this work has focused on the final co-added spectra. All the SDSS spectra are in fact taken in three or more
exposures, approximately 15 minutes long \citep{stoughton02:SDSS_EDR}, in order to facilitate cosmic ray rejection. The
exact number of exposures and the time lapsed between them varies from plate to plate, depending on the observing
conditions at the time, but exposures are usually taken back to back during the same night. Starting with DR7
\citep{abazajian09:SDSS_DR7}, the individual exposures are available separately, which opens the domain of time resolved
spectroscopy in the SDSS data base.

The goal of \sw\ is to take advantage of this situation in the context of the WD spectra, looking for RV shifts among
different exposures of the same object that may be the hallmark of a massive companion. To facilitate the search,
we have started by considering only WDs with `dominant type' DA or DB in the E06 catalog, i.e., objects that have a
relatively simple spectrum with absorption lines from either H or He. From this list, we have discarded any objects with
excess flux in the red part of the spectrum, which are usually classified as detached WD+M star binaries
\citep{silvestri06:WD+M}. This initial triage yields 7956 objects, which we have examined for RV shifts using several
automated techniques supplemented by visual examination. A detailed description of our search strategy and the
implications for completeness, the derivation of merger rates, and other related issues will be the subject of a
forthcoming publication.

\subsection{\sw\ vs. SPY}

Before going on to discuss specific results, it is instructive to outline the fundamental differences between \sw\ and
the SPY survey \citep{napiwotzki01:SPY_survey}. At a very basic level, both surveys are complementary, because the SPY
objects were examined for RV shifts from the VLT telescopes in Paranal (Chile) and have therefore mostly southern
declinations, while SDSS data are taken from the Apache Point Observatory (APO) in New Mexico and have mostly northern
declinations. The SPY survey achieved an excellent RV accuracy of $\sim2\,\mathrm{km\,s^{-1}}$ by using the
high-resolution UVES spectrograph ($R=18500$), but the demands of high-resolution spectroscopy imposed a stringent
magnitude cutoff of $B\leq16.5$ on the SPY WDs, even for the large collecting area (8.2 m) of the VLT Kueyen telescope
\citep{napiwotzki01:SPY_survey}. By comparison, the resolution that can be achieved by \sw\ with the spectrographs on
the SDSS 2.5m telescope \citep[$R=1800$,][]{york00:SDSS_Technical} is much lower, around $170\,\mathrm{km\,s^{-1}}$. But
RV resolution is not the most important factor to find potential DDWD SN Ia progenitors. The most interesting systems
are those with relatively massive components, or tight orbits, or both, which are expected to have RV shifts higher than
$\sim150\,\mathrm{km\,s^{-1}}$ \citep{napiwotzki04:SPY_04}. Moreover, by foregoing the need for high-resolution
spectroscopy we can examine dimmer WDs and hence many more objects. Through a combination of centroid fits and
cross-correlation techniques, we have been able to achieve an accuracy of $\sim120\,\mathrm{km\,s^{-1}}$, detecting RV
shifts between consecutive SDSS exposures in a WD with a $g$ magnitude of $18.9$ \citep{mullally09:DDWDs}.  Thus, \sw\
can examine roughly an order of magnitude more objects than SPY, albeit at a much lower resolution.  We can summarize
our comparison by saying that SPY is better suited for a systematic study of the properties of DDWD systems, but \sw\ is
better positioned to identify interesting objects, particularly the pre-mergers and SN Ia progenitors.

\subsection{The first \sw\ object: \WDT}

\begin{figure}
  
  \centering
  
  \includegraphics[scale=0.8,clip,angle=90]{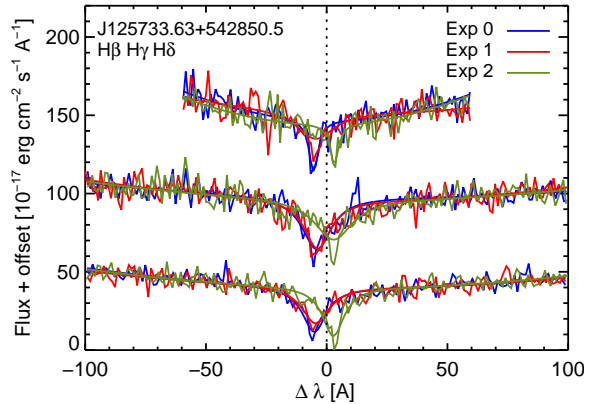}
  
  \caption{The three individual exposures of \WDT\ from the SDSS DR7 archive around the H$\beta$, H$\gamma$ and
    H$\delta$ lines. \label{fig-1}}
  
\end{figure}

To illustrate the kind of objects that can be found by \sw, we present here results for \WDT, the first CWDB identified
by the survey. In the E06 catalog, \WDT\ was listed as a DA WD with a $g$ magnitude of $16.8$ (see Table
\ref{tab-1}). There are three individual exposures of this WD in the DR7 SDSS data base.  The WD is identified as a
binary by the large shift in the centroid of the H lines between exposures 0 and 1 (taken on the night of October 3
2003) and exposure 2 (taken on the following night; see Figure \ref{fig-1}). The shift has a value of
$7.9\,\mathrm{\AA}$ at the H$_{\beta}$ line, corresponding to a RV difference of $487\,\mathrm{km\,s^{-1}}$.

\begin{deluxetable}{lc}
  \tablewidth{7.5cm}
  \tablecaption{Observational properties of \WDT \label{tab-1}}
  \tablecolumns{2}
  \tablehead{
    \colhead{Parameter} &
    \colhead{Value}
    }
    \startdata
    \cutinhead{Parameters from SDSS}
    $RA$ (J2000) & $12$h $57$m $33.63$s\\
    $Dec$ (J2000) & $+54^\circ$ $28$' $50.5$'' \\
    $u\, \mathrm{(Mag)}$ & $17.355\pm0.009$ \tablenotemark{a}\\
    $g\, \mathrm{(Mag)}$ & $16.844\pm0.004$ \tablenotemark{a}\\
    $r\, \mathrm{(Mag)}$ & $16.732\pm0.005$ \tablenotemark{a}\\
    $i\, \mathrm{(Mag)}$ & $16.722\pm0.005$ \tablenotemark{a}\\
    $z\, \mathrm{(Mag)}$ & $16.762\pm0.010$ \tablenotemark{a}\\
    Spectral type & DA \tablenotemark{b}\\
    $T_{eff}\,\mathrm{(K)}$  & $8750\pm25$ \tablenotemark{b} \\
    $log(g)$ & $9.0\pm0.005$ \tablenotemark{b}\\
    \cutinhead{Orbital parameters (APO)}
    $K_{A}\,\mathrm{(km\,s^{-1})}$ & $322.7 \pm 6.3$\\
    $P\,\mathrm{(hr)}$ & $4.5550 \pm 0.0007$ \\
    $T_{0} \mathrm{(JD)}$ & $2454868.1182 \pm 0.0006$ \\
    $\gamma_{A}\,\mathrm{(km\,s^{-1})}$ & $-28.9\pm4.6$ \\
    $\chi^2$ & $44.4$ \tablenotemark{c}\\
    \cutinhead{Spectral and Photometric Parameters (APO, SDSS)}
    $T_{eff}\,\mathrm{(K)}$ & $\sim 9000$ \tablenotemark{d}\\
    $log(g)$ & $\sim 8.5$ \tablenotemark{d}\\
    $M_{A}\,\mathrm{(M_{\odot})}$ & $0.92^{+0.28}_{-0.32}$ \tablenotemark{e}\\
    $D$ (pc) & $48^{+10}_{-19}$ \tablenotemark{e}\\
    $t_{Cool}$ (Gyr) & $2.0\pm1.0$ \tablenotemark{e}\\
    \cutinhead{Companion Parameters}
    $M_{B}sin(i)\,\mathrm{(M_{\odot})}$ & $1.62^{+0.20}_{-0.25}$\\
    $M_{B}$ for $i=60^\circ\,\mathrm{(M_{\odot})}$ & $2.10^{+0.13}_{-0.30}$\\
    $t_{Merge}$ for $i=90^\circ$ (Myr) & $511^{+342}_{-141}$\\
    $t_{Merge}$ for $i=60^\circ$ (Myr) & $267^{+165}_{-70}$ 
    \enddata

    \tablenotetext{a}{DR7 values.}
    \tablenotetext{b}{Listed {\tt autofit} results from E06.}
    \tablenotetext{c}{For 23 data points (19 d.o.f).}
    \tablenotetext{d}{Spectral parameters from the APO fits are listed without tolerance ranges due to the low quality of the fits, see Section \ref{subsec:spectrum}.}
    \tablenotetext{e}{See Section \ref{subsec:spectrum} for an explanation of the tolerance ranges on $M_{A}$ and $D$.}

\end{deluxetable}

\section{FOLLOWUP OBSERVATIONS OF \WDT}
\label{sec:Followup}

\subsection{Data acquisition and reduction}

After a WD in the SDSS catalog is identified by \sw\ as a member of a binary system, follow-up observations are
necessary to determine its period and amplitude, and obtain high signal-to-noise spectra that can be used to measure its
effective temperature $T_{eff}$ and gravity $\log{g}$, calculate its mass, and constrain the nature of its companion. In
the case of \WDT, the follow-up observations were conducted with the Dual Imaging Spectrograph (DIS) at the 3.5m ARC
telescope at APO. A total of 23 spectra were taken on the nights of February 5, 6, and 14, 2009.  We used the B1200
grating with a 1.5" slit to achieve a 0.62 \AA\ dispersion and a FWHM 1.8 \AA\ resolution. The integration time was 10
minutes for all spectra, bracketed by exposures of He, Ne and Ar arc lamps to ensure a proper wavelength
calibration. Flux was calibrated each night by taking several exposures of the spectrophotometric standard Feige~67
\citep{oke90:spectrophotometric_standards}. The data were reduced using standard long-slit IRAF routines
\citep{tody93:IRAF}. A prominent Hg emission line at 4358 \AA, courtesy of the inhabitants of White Sands, NM, confirmed
that the wavelengths are calibrated to better than the instrumental resolution: the standard deviation of the line over
8 spectra (one night) is 0.1 \AA. The final spectra span between 3770 and 5030 \AA, and clearly show all the lines in
the Balmer series from H$_{\beta}$ to H$_{10}$.

\subsection{Orbit}
\label{subsec:orbit}

\begin{figure}
  
  \centering
  
  \includegraphics[scale=0.8,clip,angle=90]{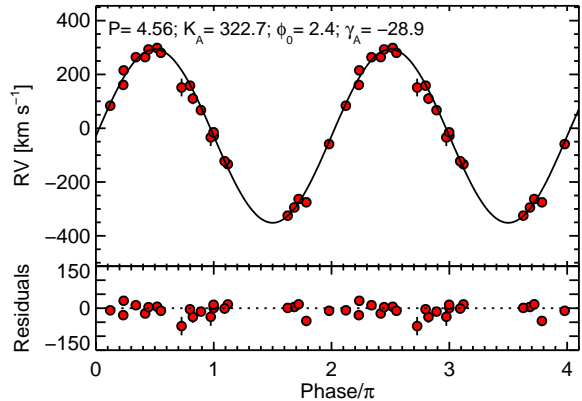}
  
  \caption{RV curve for \WDT, folded in phase with the best-fit period. \label{fig-2}}
  
\end{figure}

\begin{figure}
  
  \centering
  
  \includegraphics[scale=0.8,clip,angle=90]{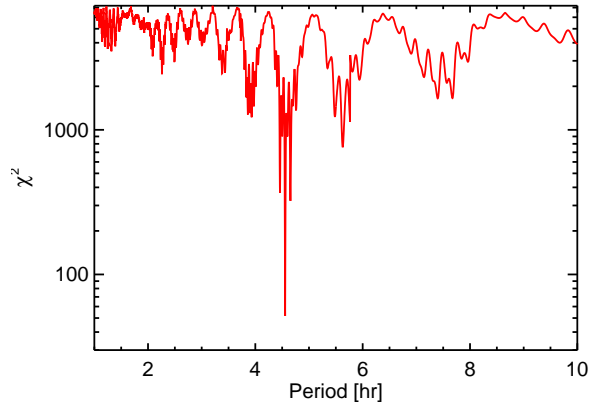}
  
  \caption{Goodness-of-fit parameter ($\chi^{2}$) as a function of period for the RV curve of \WDT. \label{fig-3}}
  
\end{figure}

We measured the RV shift of each APO spectrum by fitting the centroid of the H$_{\beta}$ line after applying the
standard Solar System barycentric corrections. We used a Lorentzian profile with a Gaussian core
\citep{thompson04:PY_Vul}, adjusting the fit parameters with MPFIT, a version of the Levenberg-Marquardt algorithm
adapted for IDL \citep{markwardt09:MPFIT}.  The resulting RV curve is well fit by a circular orbit of the form
$K_{A}\sin((2 \pi (t-T_{0})/P))+\gamma_{A}$, where $P$ is the period, $K_{A}$ the semiamplitude, $T_{0}$ a temporal
offset marking the zero-point of the curve, and $\gamma_{A}$ a constant RV offset (see Figure \ref{fig-2}). We found
that the nine day baseline of the APO spectra strongly constrains the period of the binary to $P=4.5550 \pm 0.0007$ hr,
(Figure \ref{fig-3}), with a semiamplitude of $K_{A}=322.7 \pm 6.3\,\mathrm{km\,s^{-1}}$ (see Table \ref{tab-1} for a
complete list of the orbital parameters).

\subsection{Spectrum}
\label{subsec:spectrum}

\subsubsection{Spectral fits}

The {\tt autofit} results from E06 give $T_{eff}=8750\pm25$ K and $\log{g}=9.0\pm0.005$ for \WDT, but these parameters
are derived from the co-added DR4 SDSS spectrum of an object that was not known to be a short-period binary. In order to
make the best possible determination of $T_{eff}$ and $\log{g}$, we have co-added the 23 APO spectra after removing the
RV shifts. The resulting spectrum has a S/N of $160$, much higher than the co-added SDSS spectrum (S/N=$30$). 

\begin{figure*}
  
  \centering
  
  \includegraphics[scale=0.8,clip,angle=90]{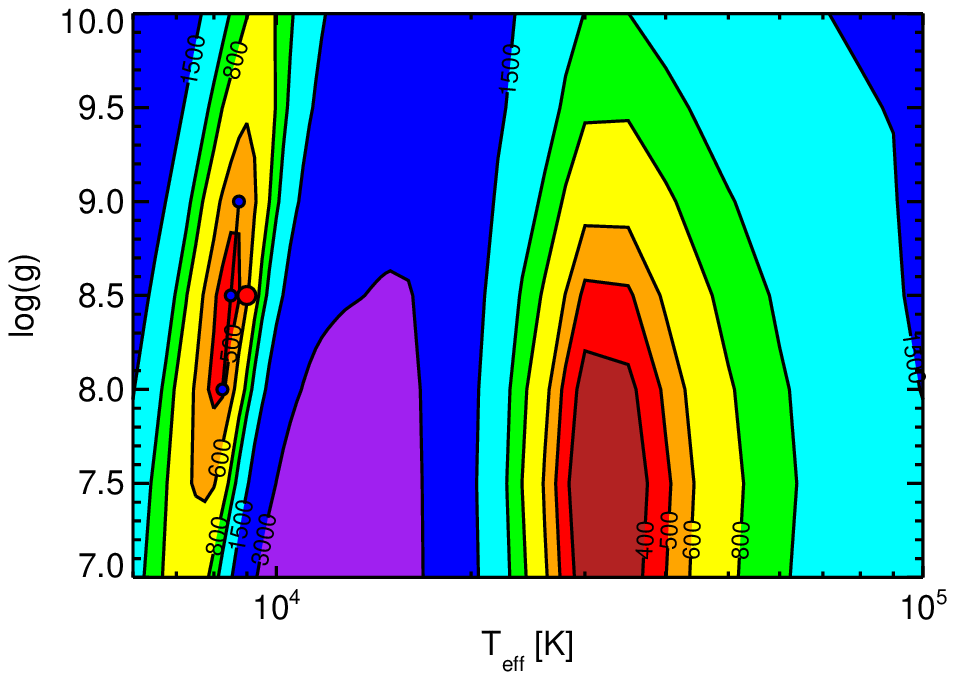}
  \includegraphics[scale=0.8,clip,angle=90]{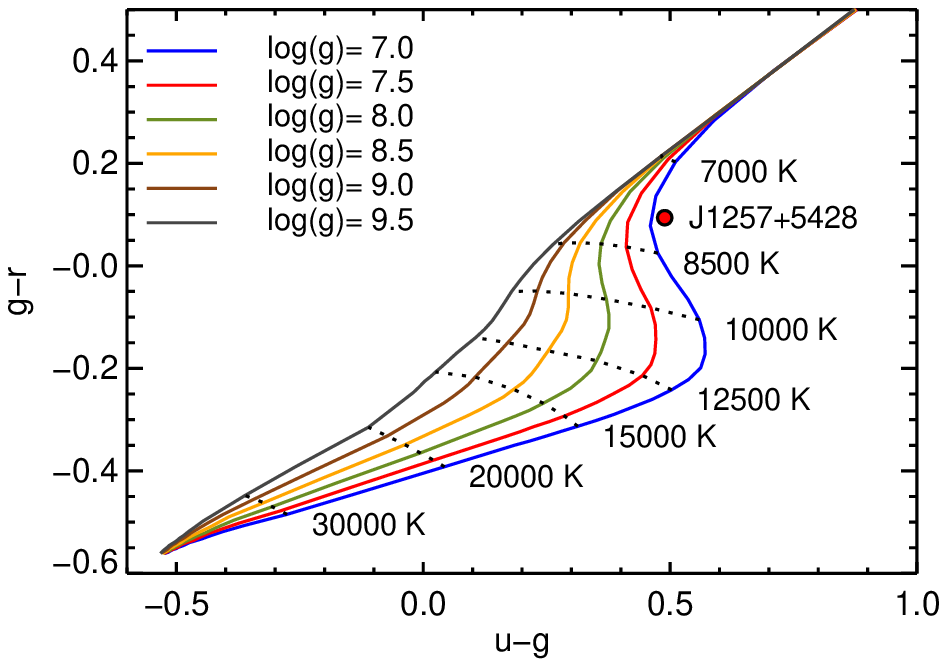}
  
  \caption{Constraints on $T_{eff}$ and $\log{g}$ for \WDT. Left: surface contours of the raw $\chi^{2}$ parameter
    obtained from fitting the co-added APO spectrum of \WDT\ to the most recent version of the models by
    \citet{finley97:WD_models_spectral}. The best-fit solution ($T_{eff}=9000$ K, $\log{g}=8.5$, see text and Figure
    \ref{fig-5}) is indicated by a large red circle. The models used to derive the confidence ranges for the WD mass are
    indicated by smaller blue circles, joined by straight lines. Right: comparison between the DR7 SDSS photometry of
    \WDT\ (red circle) and the photometric models of \citet{holberg06:WD_models_photometry}. The cooling tracks are
    labeled by $\log{g}$, with points at the same $T_{eff}$ represented by dotted lines. \label{fig-4}}
  
\end{figure*}

\begin{figure*}
  
  \centering
  
  \includegraphics[scale=0.8,clip,angle=90]{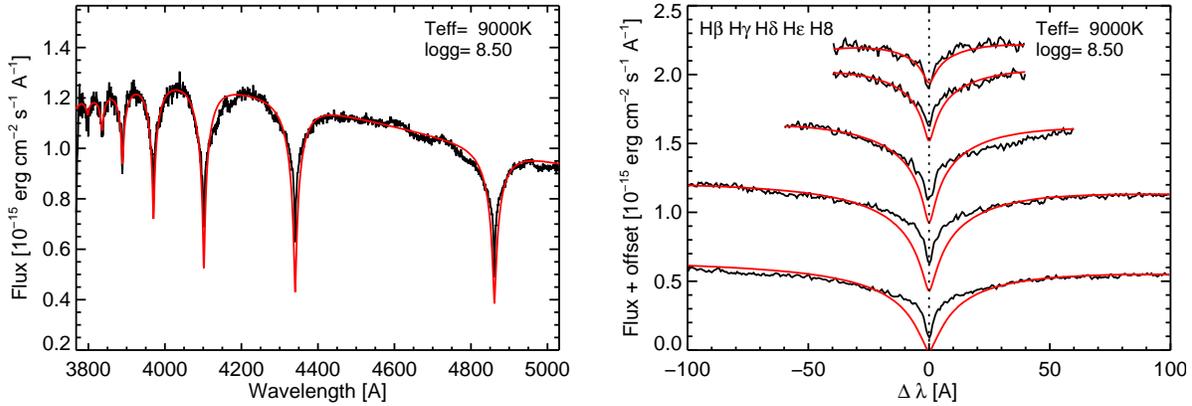}

  \caption{Best-fit spectral solution for \WDT: $T_{eff}=9000$ K, $\log{g}=8.5$.\label{fig-5}}
  
\end{figure*}

We have used the MPFIT routine to fit the co-added APO spectrum with the latest (June 2008) version of the Koester
atmosphere models descrbed in \citet{finley97:WD_models_spectral}, adjusted with a first order polynomial. This spectral
fitting has proved a challenging endeavor. The raw (unreduced) $\chi^{2}$ parameter has two distinct minima in the
$T_{eff}$, $\log{g}$ plane: one around $T_{eff}=35000$ K, $\log{g}=7.5$ and one around $T_{eff}=8500$ K, $\log{g}=8.5$
(see Figure \ref{fig-4}, left panel). The hot solutions provide slightly lower $\chi^{2}$ values than the cold
solutions, but neither fits are very impressive (see Figure \ref{fig-5} for the best-fit cold solution - the visual
quality of the best-fit hot solution is similar). This degeneracy between cold and hot spectral solutions is common in
DA WDs, and it can usually be broken by including photometric information \citep{kleinman04:SDSS_WDs}. In the right
panel of Figure \ref{fig-4}, we compare the DR7 SDSS photometry of \WDT\ (Table \ref{tab-1}) with the photometric models
of \citet{holberg06:WD_models_photometry}.  The hot solutions can be discarded with a high degree of confidence. The
cold solutions come very close to reproducing the SDSS photometry, with the models along the cold $\chi^2$ minimum being
offset from the data by only $\sim -0.15$ magnitudes in $u-g$. This kind of discrepancy is not uncommon in comparisons
between the SDSS photometry of DA WDs and cooling models \citep[][see also Section \ref{subsubsec:poorfits}
below]{holberg06:WD_models_photometry}. Including the full reddening correction of \citet{schlegel98:reddening} changes
the $u-g$ color by $-0.025$ magnitudes and the $g-r$ color by $-0.018$ magnitudes, which reduces the discrepancy by a
small amount.

Even though it is not exactly at the bottom of the local $\chi^{2}$ minimum, we have found that the spectral model with
$T_{eff}=9000$ K and $\log{g}=8.5$ (Figure \ref{fig-5}) provides the best match to the continuum and the cores of the
high order Balmer lines, which should be more sensitive to the value of $\log{g}$
\citep[][]{finley97:WD_models_spectral}. We consider this model our `best fit', although it cannot be said that the
spectral adjustment shown in Figure \ref{fig-5} is particularly good. Nevertheless, the values of $T_{eff}$ and
$\log{g}$ that we obtain are reasonably close to those found by E06 using the co-added SDSS spectrum. This agreement
between different analysis techniques applied to different data sets of the same object is encouraging, and supports our
choice of spectral model.

\subsubsection{Why are the fits so poor?}
\label{subsubsec:poorfits}

The failure of theoretical models to reproduce the details of the spectrum and photometry of \WDT\ is not entirely
unexpected. Even the most updated versions of the Koester atmosphere models have problems for WDs with $T_{eff}<12000$
K. At these temperatures, the contribution of neutral and charged perturbers to the equation of state is of the same
order of magnitude, and the size of the convective envelope of the WD has an abrupt increase
\citep{koester08:WDs_low_Teff}. These effects have been put forward as possible causes for the observed increase of the
average effective gravity for WDs in this temperature range
\citep{kepler07:WD_mass_distribution,bergeron07:mass_distribution_WDS}. The discrepancies shown in Figure \ref{fig-5},
however, are rather large, and they should not be written off as shortcomings of the models before considering
alternative explanations.

The high S/N of our co-added APO spectrum reveals some unusual features in the continuum around 4600 \AA\ and 4750
\AA. The features are not present in the photometric standard Feige 67, which indicates that the flux calibration is not
at fault. None of the calibration frames, flat fields or biases have flaws that could have caused these features. Other
WDs that we observed with the same spectrograph in the same mode \citep{mullally09:DDWDs} do not show them, but these
are fainter objects with worse S/N in the co-added spectra. At this stage, we have no explanation for the presence of
these localized features, but they are clearly not large enough to influence the mismatch between the models and the
observed spectrum.

\begin{figure}
  
  \centering
  
  \includegraphics[scale=0.8,clip,angle=90]{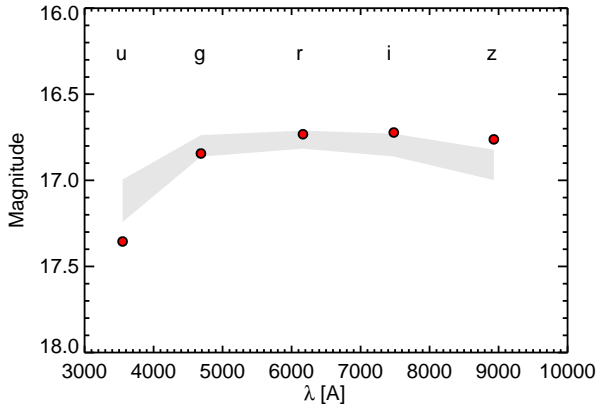}

  \caption{Comparison between the SED of \WDT\ from the five-band DR7 SDSS photometry (red circles) with the predictions
    of \citet{holberg06:WD_models_photometry} for WD models along the cold $\chi^2$ minimum, $T_{eff}$ between 8000 and
    9000 K and $\log{g}$ between 8.0 and 9.0 (shaded area). \label{fig-6}}
  
\end{figure}

It is possible that the spectrum of \WDT\ is simply distorted by optical emission from its binary companion. The
co-added SDSS spectrum shows no traces of line emission or a substantial excess in the red spectrograph, clear signs of
the presence of a nondegenerate companion that would have prompted a DA+M classification for \WDT\ by E06 (this is ruled
out by the RV curve as well, see the discussion in Section \ref{sec:Companion}). A WD companion is more difficult to
discard from the point of view of the spectrum alone.  We have examined the individual APO spectra that are closer to
quadrature, when the relative RV shift of the individual components should be larger, but found no sign of artifacts or
asymmetries in the Balmer lines that might indicate the presence of a second DA WD of similar temperature. A
substantially hotter WD of any spectral class would rapidly overwhelm the spectrum of \WDT, having a large impact on the
SDSS photometry shown in Figure \ref{fig-4}. The only possibility left is a non-DA WD of similar or lower temperature
with a rather featureless spectrum. In Figure \ref{fig-6} we display the spectral energy density (SED) of \WDT\,
together with the range covered by the models of \citet{holberg06:WD_models_photometry}. The SEDs of models along the
cold $\chi^2$ minimum are roughly consistent with the five-band photometry of \WDT. There might be a hint of an excess
in the $z$ band, but infrared data would be required to confirm this possibility. On the other hand, all the models seem
to overpredict the flux in the $u$ band, an effect that was already apparent in the color-color plot shown in Figure
\ref{fig-4}. Since this is a flux deficit instead of an excess, it is hard to explain by invoking the presence of a
companion. In fact, \citet{holberg06:WD_models_photometry} find similar problems in the $u$ band for other cold DA WDs
(see their Figure 6).

We conclude that the shortcomings of the WD models in this temperature range are indeed the most likely explanation for
the poor quality of the spectral fit shown in Figure \ref{fig-5}. A cold WD companion cannot be ruled out based on the
spectral information alone, but neither is there any evidence in the observations that \textit{requires} the presence of
such a companion. We shall revisit this possibility in Section \ref{sec:Companion}.

\subsubsection{Estimating the mass of the primary WD}

The issues with the models that we have discussed require that we be extremely careful with the interpretation of the
spectral fits.  In particular, we cannot justify the calculation of confidence ranges around the fitted parameters in
the usual way (i.e., based on the $\chi^{2}$ statistic). But we are not interested in the confidence ranges around
$T_{eff}$ and $\log{g}$ \textit{per se}. Rather, our goal is to use the spectrum of \WDT\ to measure its mass and
constrain the nature of its unseen companion.

\begin{figure}
  
  \centering
  
  \includegraphics[scale=0.8,clip,angle=90]{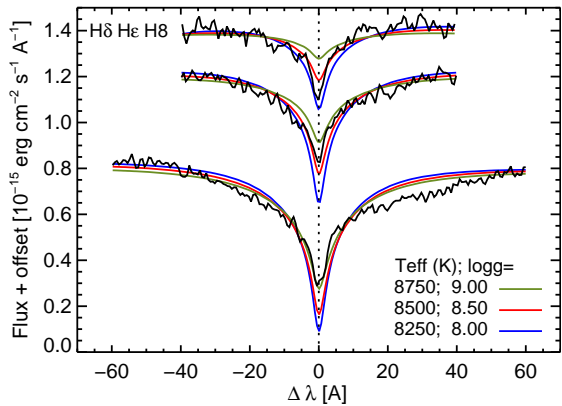}

  \caption{Comparison between the high order Balmer lines (H$_{\delta}$, H$_{\epsilon}$, and H$_{8}$) in the spectrum of
    \WDT\ and the three models along the cold $\chi^{2}$ minimum indicated with small blue circles in the left panel of
    Figure \ref{fig-4}. Although higher order lines (H$_{9}$ and H$_{10}$) are present in the spectrum, the S/N is too
    low for a meaningful comparison with the models. \label{fig-7}}
  
\end{figure}

An estimate for the mass of \WDT\ can be obtained by examining the performance of the spectral models around the cold
minimum in $\chi^{2}$ space. The cooling curves of \citet{fontaine01:WD_Cosmochronology} included in the models of
\citet{holberg06:WD_models_photometry} give a value of $M_{A}=0.92\,\mathrm{M_{\odot}}$ for our best-fit spectral
solution, which is probably a good initial guess for the mass of \WDT. Along the cold $\chi^{2}$ minimum, the spectral
solutions with $T_{eff}=8250$, $8500$, and $8750$ K; and $\log{g}=8.0$, $8.5$, and $9.0$ (indicated by the small blue
circles in the left panel of Figure \ref{fig-4}) appear to bracket the effective gravity values that are in best
agreement with the higher order Balmer lines in the spectrum, as shown in Figure \ref{fig-7}. Any models with $\log{g}$
outside this range result in extremely poor fits and must be discarded, including the models with lower $T_{eff}$ and
lower $\log{g}$ that seem to be preferred by the photometry (right panel of Figure \ref{fig-4}). Since we have already
ruled out a DA WD companion of similar temperature, we expect the shape of the higher order Balmer lines to be a
reliable probe of the mass of \WDT, even if the underlying continuum is somewhat distorted.

The spectral models shown in Figure \ref{fig-7} translate into masses between $0.60$ and $1.20\,\mathrm{M_{\odot}}$,
which we will adopt as conservative lower and upper limits to $M_{A}$. It is unfortunate that the poor performance of
the spectral models in this temperature range does not allow for a more accurate measurement of $M_{A}$, but we believe
that our estimate is reasonable, and probably the best that can be done with current tools. We note that the models we
contemplate here span a range of $\log{g}$ values around the best-fit solution ($\pm 0.5$) that is comfortably larger
than the reported increase for $T_{eff}<12000$ K \citep[$\sim
0.2$,][]{kepler07:WD_mass_distribution,bergeron07:mass_distribution_WDS}. Also, our lower limit to $M_{A}$ is close to
the mean mass of DA WDs with $T_{eff} \geq 12000$ K
\citep[$0.58\,\mathrm{M_{\odot}}$,][]{kepler07:WD_mass_distribution}, which would be the best guess for the mass of a WD
in absence of any spectral or photometric information. For all these reasons, we are confident that the true mass of
\WDT\ lies within the range of values that we propose here. For comparison purposes, the best-fit hot spectral model
($T_{eff}=35000$ K, $\log{g}=7.5$) translates into a value of $0.45\,\mathrm{M_{\odot}}$ for $M_{A}$. With the set of
limiting spectral models around the best-fit cold solution shown in Figure \ref{fig-4}, the
\citeauthor{holberg06:WD_models_photometry} curves yield a cooling age of $2.0\pm1.0$ Gyr. The absolute $g$ magnitude is
$13.4^{+1.1}_{-0.4}$, which results in a distance of $D=48^{+10}_{-19}$ pc for \WDT.

\section{THE COMPANION OF \WDT: NEUTRON STAR OR BLACK HOLE?}
\label{sec:Companion}

\begin{figure}
  
  \centering
  
  \includegraphics[scale=0.8,clip,angle=90]{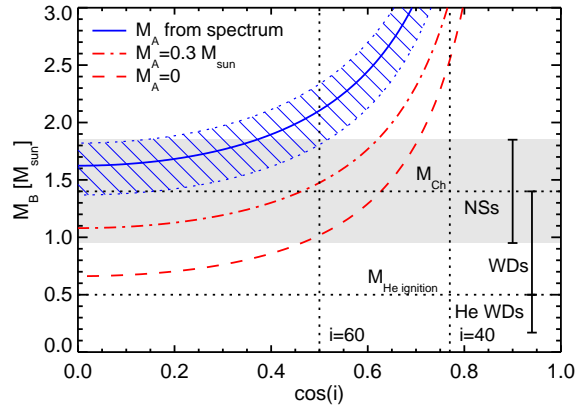}

  \caption{Mass function for the companion of \WDT. The solid blue curve represents the values obtained using the best
    estimate for the WD mass ($M_{A}=0.92\,\mathrm{M_{\odot}}$), and the striped region is bound by the upper and lower
    limits to the WD mass obtained from the spectrum ($M_{A}=0.60$ and $1.20\,\mathrm{M_{\odot}}$). For comparison, the
    mass functions for $M_{A}=0.3\,\mathrm{M_{\odot}}$ and $M_{A}=0$ are also shown (dash-dotted and dashed red
    plots). The range of masses for known NSs from \citet{lattimer07:NS_EOS} has been highlighted with a shaded region,
    taking into account the recent re-evaluation of the mass of PSR 0751+1807 by
    \citet{nice08:no_massive_NS}. \label{fig-8}}
  
\end{figure}

Combining our estimate for the mass of \WDT\ ($M_{A}=0.92^{+0.28}_{-0.32}\,\mathrm{M_{\odot}}$) with the values of $P$
($4.5550 \pm 0.0007$ hr) and $K_{A}$ ($322.7 \pm 6.3\,\mathrm{km\,s^{-1}}$) derived in Section \ref{subsec:orbit}, we
obtain $M_{B}\sin(i)=1.62^{+0.20}_{-0.25}\,\mathrm{M_{\odot}}$ for its unseen companion, with $M_{B}$ and the
inclination angle $i$ being degenerate as in all single-lined binaries. A plot of the companion mass as a function of
$\cos(i)$ is shown in Figure \ref{fig-8}.  We stress that the properties of the RV curve alone require a massive
companion, regardless of the estimated value for $M_{A}$: $M_{B}$ must be more massive than $0.66 \,
\mathrm{M_{\odot}}$ even for $M_{A}=0$ (i.e., assuming the value of $M_{A}$ is negligible for the dynamics of the
system), and more massive than $1.08 \, \mathrm{M_{\odot}}$ for $M_{A}=0.3 \, \mathrm{M_{\odot}}$. With any reasonable
range of masses for the primary, a nondegenerate stellar companion would have a spectral type G or earlier, which is
clearly incompatible with the observations of \WDT. The companion must therefore be a compact object, either another WD,
a NS or a BH.

Could the companion be a WD? We have not been able to discard this possibility from the point of view of the optical
spectrum alone, but the properties of the RV curve make it very unlikely. The companion could only be a WD if our mass
estimate for $M_{A}$ was completely off, and even then, it would have to be a very massive object (above $1.0 \,
\mathrm{M_{\odot}}$ for any reasonable value of $M_{A}$), cool object, without any prominent absorption lines (see
discussion in Section \ref{subsubsec:poorfits}). Such massive WDs are extremely rare. But we have found no evidence to
indicate that the shape of the Balmer lines in the spectrum of \WDT\ is distorted in any way, so there is no reason to
doubt our conservative estimate for $M_{A}$. If this estimate holds, assuming the smallest possible value of $M_{A}$
($0.6\,\mathrm{M_{\odot}}$), the orbit would have to be nearly edge-on ($i \geq 82^\circ$) for $M_{B}$ to be below the
Chandrasekhar limit $M_{Ch}$. This inclination is expected to happen randomly in only $14\%$ of binary systems, but \sw\
does have a strong observational bias towards finding high inclination systems. Even then, the companion would be more
massive than the largest known WDs \citep[$1.33$ to $1.35\,\mathrm{M_{\odot}}$, see][]{kepler07:WD_mass_distribution,
  barstow95:massive_WD}. In this unlikely circumstance, the \WDT\ system would have a combined mass at least $43\%$
larger than $M_{Ch}$, but might not be a SN Ia progenitor if the companion turns out to be an ONe WD instead of a CO WD
\citep{garcia-berro97:stars}. Considering all the evidence, the most likely possibility by far is that the companion of
\WDT\ be the stellar remnant of a supernova explosion, either a NS or a BH. We note that our estimated distance range
places this object closer to the Solar System than any other known NS \citep{posselt07:M7}.

The puzzle of the nature of the companion of \WDT\ cannot be solved with the observations that we present in this paper,
and must be the subject of future work. It is, however, interesting to speculate about the possibilities. For our
best-fit value of $M_{A}$ ($0.92\,\mathrm{M_{\odot}}$), $M_{B}$ is below $1.86\,\mathrm{M_{\odot}}$ \citep[the largest
masured mass for a NS, see][]{lattimer07:NS_EOS,nice08:no_massive_NS} for $i \geq 67^{\circ}$, which has a random
likelihood of $39\%$. In this case, the mass of the WD and the circularity of the orbit would place the system in the
class of `intermediate-mass' WD+NS binaries \citep[see Table 1 in][]{stairs04:pulsars_binary_systems}. Current models
for binary stellar evolution predict that these systems undergo unstable mass transfer during a common-envelope phase
\citep{stairs04:pulsars_binary_systems}, which results in the NS becoming a mildly recycled millisecond pulsar
(MSP). This scenario would explain the high inferred mass for the companion of \WDT, well above the average for a NS, as
a byproduct of the mass transfer process \citep{lattimer07:NS_EOS}. If the NS is indeed an MSP, we expect the magnetic
field to be low ($\sim10^{9}$ Gauss), the pulsar lifetime to be large ($\sim1$ Gyr), and the opening angle to be wide
\citep[possibly many tens of degrees, see][]{phinney94:binary_and_ms_pulsars}. Under these conditions, the prospects for
detecting such a nearby MSP are good, although we note that the cooling time for the WD is also of the order of Gyr (see
Section \ref{subsec:spectrum}), and the MSP might have lost a large part of its magnetic field. For
$M_{A}=0.92\,\mathrm{M_{\odot}}$ and inclination angles below $67^{\circ}$ ($61\%$ random likelihood), the companion
would probably be a stellar mass BH. In this case, we would not expect any kind of direct emission from it.

No radio or X-ray source appears at this location in any of the indexed astronomical catalogs, but this does not
preclude the existence of a faint counterpart to the companion of \WDT. This part of the sky has only been shallowly
surveyed for MSPs in the radio. The most stringent limits are probably from the Green Bank 140 ft 350 MHz survey, which
had a nominal sensitivity of 12-15 mJy \citep{sayer97:GBT_northern_sky_pulsar_survey}. In the X-rays, this location has
never been observed with \textit{Chandra} or \textit{XMM-Newton}, and a faint nearby source could have easily escaped
detection by the \textit{ROSAT} All-Sky Survey. A systematic cross-correlation between stellar sources in SDSS
(including all the WDs from E06) and the \textit{ROSAT} catalogues was performed by \citet{agueros09:ROSAT_SDSS_Stars},
who found no counterpart to \WDT.

One interesting implication of the companion of \WDT\ being a NS or BH is that the system should have received some kind
of kick from the SN explosion. In principle, the spatial velocity of \WDT\ can be determined by the temporal average of
the non-gravitational Doppler shifts in the spectrum, which gives the radial component, and the proper motion measured
by SDSS ($\mu=0.049\,\mathrm{mas\,yr^{-1}}$), which gives the component on the plane of the sky. Taking our distance
estimate, the proper motion translates into a transverse velocity of $11^{+2}_{-4}\,\mathrm{km\,s^{-1}}$. The radial
component is more difficult to estimate, because the constant component to the RV curve
($\gamma_{A}=-28.9\pm4.6\,\mathrm{km\,s^{-1}}$ in our fit, see Section \ref{subsec:orbit}) is the combination of the
true radial velocity and the gravitational redshift of the WD. The value of the gravitational redshift depends on the WD
mass, which we cannot measure with accuracy (see Section \ref{subsec:spectrum}), and the WD radius, which is also model
dependent. A detailed estimate of the gravitational redshift for \WDT\ is outside the scope of this work, but for a
massive $\sim1\,\mathrm{M_{\odot}}$ WD, we expect it to be of the order of $90\,\mathrm{km\,s^{-1}}$
\citep{wegner91:gravitational_redshift_WDs}, which would require a radial velocity around
$-120\,\mathrm{km\,s^{-1}}$. The total spatial velocity would then be $\sim120\,\mathrm{km\,s^{-1}}$, mostly in the
radial direction, which is comparable to the measured kicks for pulsars in binary systems \citep{wang06:NS_kicks}.

Regardless of what the nature of the companion to \WDT\ turns out to be, the system is clearly very interesting from an
astrophysical point of view, and new observations should yield exciting results in the near future. From the measured
orbital parameters, the separation of the components must be small, with values of the semimajor axis larger than $0.02$
AU only for $i \leq 17^\circ$ ($0.0093$ AU at $i = 60^\circ$). If pulsations from a NS companion were detected, this
could allow for a significant measurement of the Shapiro delay, as in PSR J0437$-$4715
\citep{vanstraten01:Shapiro_delay}. The merging time $t_{Merge}$ of the \WDT\ system due to GW radiation is $\leq
511^{+342}_{-141}$ Myr , with the uncertainty in this upper limit arising from the uncertainty in the determination of
$M_{A}$. For the canonical inclination angle $i = 60^\circ$, $t_{Merge}$ becomes $267^{+165}_{-70}$ Myr. We note that
only three WD+NS binaries with $t_{Merge} < t_{Hubble}$ were known previously
\citep{stairs04:pulsars_binary_systems,kim04:Ns_WD_Mergers}, and \WDT\ has a shorter period and $t_{Merge}$ than any of
them. The discovery of a fourth object in this class should prompt a revision of the estimated WD+NS merger rates and
their expected contribution to the GW background \citep{kim04:Ns_WD_Mergers}. 

The ultimate fate of \WDT\ is unclear. \citet{king07:WD+NS_GRBs} have proposed the merging of massive WDs onto NSs as a
scenario for the origin of long-duration gamma-ray bursts without an accompanying SN like GRB 060614
\citep[][]{gal-yam06:GRB060614,dellavalle06:GRB060614}, but theoretical simulations for the final accretion phase in
this kind of events have not been performed yet. The coalescence of a BH with a massive WD may also lead to a gamma-ray
burst \citep{fryer99:WD_BH_Mergers}, but the observational signature depends again on the details of the final accretion
phase, which are poorly understood. More exotic astrophysical transients, such as may be observed by present
\citep[PTF,][]{rau09:PTF_science} and forthcoming \citep[e.g. LSST,][]{ivezic08:LSST} synoptic surveys, cannot be
discarded as outcomes.

\section{CONCLUSIONS}
\label{sec:Conclusions}

In this paper and a companion publication \citep{mullally09:DDWDs} we have presented the first results from the SWARMS
survey, an ongoing project aimed at discovering and characterizing CWDBs in general and DDWD SN Ia progenitors in
particular among the WDs in the spectroscopic SDSS data base. Further results and a complete description of the survey
will be the subject of forthcoming publications. Our ultimate goal is to estimate a rate for the merger of DDWD SN Ia
progenitors in the Galaxy, and to use that rate, in combination with the results from SPY and other surveys, to assess
the viability of the DDWD progenitor scenario for Type Ia SNe.

To demonstrate the capabilities of SWARMS, we have focused on \WDT, the first object found by the survey. This WD was
identified as a member of a binary system from the SDSS spectra, and follow-up observations were performed to measure
its orbital and spectral parameters. The RV curve has a period of $4.5550 \pm 0.0007$ hr, with a semiamplitude of $322.7
\pm 6.3\,\mathrm{km\,s^{-1}}$. The primary WD is cold ($\sim9000$ K), which makes the spectral analysis somewhat
challenging, but we have derived a conservative estimate of $0.92^{+0.28}_{-0.32}\,\mathrm{M_{\odot}}$ for its
mass. This implies that the unseen companion must be larger than $1.62^{+0.20}_{-0.25}\,\mathrm{M_{\odot}}$, and is
probably a NS or a BH. At a distance of $D=48^{+10}_{-19}$ pc, this would be the closest remnant of a SN explosion to
the Solar System. We have discussed the implications of our discovery, and suggested future avenues of research on this
object. This kind of massive, close binary is a perfect example of the objects that can be found by SWARMS and other
similar surveys like MUCHFUSS \citep{tillich09:MUCHFUSS}. We anticipate more such discoveries, which we will report in
the literature. We also plan to set up an on-line database of SWARMS objects.

\acknowledgements{We are grateful to the APO staff for assistance during our observations, to Alexei Filippenko for
  hosting CB at Berkeley during the initial stages of this project, and to the referee, Ralf Napiwotzki, for several
  comments that improved the quality of our manuscript. We also wish to thank Detlev Koester, who kindly provided an
  updated version of his grid of WD models and contributed useful suggestions. We acknowledge fruitful discussions with
  Lars Bildsten, Jenny Greene, Gijs Nelemans, Roman Rafikov, Scott Ransom, Anatoly Spitkovsky, and Michael Strauss. At
  Princeton University, CB is supported by NASA through Chandra Postdoctoral Fellowship Award Number PF6-70046 issued by
  the Chandra X-ray Observatory Center, which is operated by the Smithsonian Astrophysical Observatory for and on behalf
  of NASA under contract NAS8-03060. CB also thanks the Benoziyo Center for Astrophysics for support at the Weizmann
  Institute of Science. SET thanks the Crystal Trust for their financial support.}

\end{document}